%% file: conv_arx.tex
\documentclass[aps,pre,twocolumn,showpacs,reprint,superscriptaddress,floatfix]{revtex4-2}

\usepackage{graphicx}
\usepackage{amsmath,amssymb}
\usepackage{hyperref}
\usepackage{xcolor}

\usepackage{url}
\usepackage{hyperref}

\begin{document}

\title{Flat-histogram algorithms: optimal parameters and extended application}

\author{Timur Shakirov}
\affiliation{Institute of Physics, Martin Luther University of Halle-Wittenberg, Halle, Germany}

\date{\today}

\begin{abstract}
We provide analysis of the convergence properties and applicability extensions of flat-histogram algorithms, with a particular focus on the Wang-Landau algorithms (exemplified by converging stochastic approximation Monte Carlo (SAMC)) and multicanonical (MUCA) algorithms. Our investigation reveals that the optimal decay rate of the modification factor in SAMC algorithms is influenced by the number of energy bins rather than the width of the energy range. Despite the frequent naming of these algorithms based on the histogram flatness, our findings indicate that flatness demonstrates a limited correlation with  estimation accuracy. 
We explore the implications of integrating the importance sampling technique with flat-histogram algorithms, demonstrating that this combination yields comparable or better accuracy in density of states estimations, almost independent of specific algorithmic parameters within certain bounds. Furthermore, our research extends the possibilities of the flat-histogram and importance sampling combination for investigating a range of underlying system parameters simultaneously within a single simulation. These system parameters could both originate from the potential, e.g., various relative contributions of different energy terms or characteristic interaction range, and characterize the accessible configurations, e.g., through the size of the simulation box.

\end{abstract}

\maketitle

\section{Introduction}

Monte Carlo sampling methods based on estimates of the free energy, entropy or density of states have been developing for half a century \cite{umbrella1972,umbrella1977,muca1,
muca2,muca3,muca4,transmat1,transmat2,transmat3,wl1,
wl2,wl1t1,wl1t2,wl1t3,wl1t4} and have multiple applications varying from investigations of complex free energy landscapes \cite{ex_polymer_1,ex_polymer_2,ex_polymer_3,ex_polymer_4,
ex_polymer_5,ex_polymer_6,ex_polymer_7,ex_polymer_8,ex_polymer_9,
ex_polymer_10,ex_polymer_11,ex_polymer_12,ex_polymer_13,ex_polymer_14,
ex_2d,TSaggregation,
ex_protein_1,ex_protein_2,ex_protein_3,ex_protein_4,ex_protein_5,
ex_protein_6,ex_protein_7,ex_protein_8,
ex_ising_1,ex_ising_2,ex_ising_3,ex_ising_4,ex_ising_5,ex_ising_6,
ex_ising_7,ex_ising_8} to implementations within artificial intelligence \cite{ex_ai_1,ex_ai_2,ex_ai_3}.
These algorithms allow for sampling a broad range of energies (or other parameters of interest) during a single simulation, which is practically impossible for the conventional Metropolis algorithm \cite{metropolis1953, hastingsMC}. The modern development of new techniques offers the possibility of almost uniform sampling over a desired energy range. This feature of the algorithms motivates one of their frequent namings: flat-histogram algorithms.

Development of the simulation techniques provide a variety of algorithms and their modifications known as: entropic sampling, umbrella sampling, multicanonical sampling, Wang-Landau algorithm, stochastic approximation Monte Carlo etc. These algorithms are different in particular details of entropy or free energy approximation and its modification during a simulation run. The choised keeping the initial estimation or frequent update (e.g., after each trial move) of the estimated entropy or free energy during the simulation allows uniting a variety of algorithms into two groups: The multicanonical-like algorithms (MUCA-like) and the Wang-Landau-like (WL-like) algorithms correspondingly. Since this is the most fundamental difference between algorithms in the rest of the paper we consider properties of the two convergent archetypical algorithms: the multicanonical algorithms (MUCA) \cite{muca1,muca2,muca3} and the stochastic approximation Monte Carlo algorithms (SAMC) \cite{samc1,samc2,samc3}.

The accuracy and the convergence rate are without doubt the most important properties of algorithms. The problem of convergence is always in focus of researchers \citep{transmat2,wl1t1,wl1t2}, especially on the development stage of new types of algorithm, such as the $1/t$~Wang-Landau algorithm \cite{wl1t1,wl1t2,wl1t3,wl1t4}. While the problem of convergence is conceptually solved, for instance for SAMC and MUCA algorithms the mean-square deviation from the exact function is expected to be inversely proportional to the simulation length $\propto 1/t$ \cite{samc1,samc2,samc3,muca3,TSconvergence}. The relationship between the parameters of an algorithm and the expected accuracy has not yet been thoroughly investigated. This is especially important for WL-type algorithms, as they depend on a broader set of parameters.

An additional facet of the algorithms' efficiency is the potential to broaden data accumulation. This extends beyond a single function, which determines the trial move acceptance probabilities, to include a broader set of models or energy functions. While this possibility within a broad-histogram sampling was already discussed in the earliest works \cite{umbrella1972,umbrella1977} in terms of altering the temperature or parameters of the interaction potential, such extensions were not addressed in the literature in the context of modern flat-histogram techniques.

\section{Methods and models}
\subsection{Flat-histogram Monte Carlo methods}
Flat-histogram algorithms require the initial estimation of the probability distribution as a function of the parameter of interest, e.g., energy. Since the typical variation of the probability values covers many orders of magnitude, it is convenient to present it in logarithmic form.
We denote the estimated density of states as $g\left( E\right)$, correspondingly $\ln g\left( E\right) $ is proportional to the entropy or (generalized) free energy of the system. Note that the results described below are based on sampling of conformational energies ($E=U$), but could be applied to sampling of any parameter or their combinations ($E= \{U,a,\dots \}$ \cite{ex_protein_6,ex_2d}) or multidimensional parameters.

A general scheme of the flat-histogram algorithm could be summarized as:
\begin{enumerate}
\item[I:] Choose an initial estimate of the DOS $g\left(E\right)=g_0\left( E\right)$;
\item[II:] make a trial change of the system configuration from the current state $x_0$ to a trial state $x_1$, accept the new state with probability:
\begin{equation}
\label{eq:acc}
p_{\text{acc}} = \min\left[1; \frac{g\left(E_0\right)}{g\left(E_1\right)} \right],
\end{equation}
with $E_i = E\left(x_i\right)$;
\item[III:] if necessary, update the current estimate of the DOS;
\item[IV:] repeat steps II and III until the finalization criteria are fulfilled. 
\end{enumerate}

The choice of the initial estimate of the DOS, $g_0\left( E\right)$, strongly influences the convergence of the simulation. For a bad choice of the initial estimate, a long computational time will be required to compensate the initial inaccuracy, and the efficiency of the method decreases. A reasonable way to avoid this problem is using the final estimation of the DOS obtained in a previous simulation of the system. An alternative estimation could be provided in a series of few short preliminary runs. The goal of the preliminary run or runs is not to reach a well-convergent DOS, but get only a rough estimate, which provides a better starting point than a trivial uniform distribution.

While steps I and II are similar in realization for all flat histogram methods, the realization of step III depends strongly on the chosen algorithm. For the MUCA sampling the step III occurs only once at the end of the simulation. The new estimate of the DOS is calculated as:
\begin{equation}
\label{eq:muca}
\ln g\left(E\right) \rightarrow \ln g\left(E\right) + \ln {H\left(E\right)} 
\end{equation}
with $H\left(E\right)$ being the visitation histogram, i.e., the counter of visits to the parameter value $E$ (or corresponding bin for the continuous parameters) during the simulation.

For the SAMC algorithms, the DOS is updated after each trial move as follows
\begin{equation}
\label{eq:samc}
\ln g\left(E;t+1\right) \rightarrow \ln g \left(E;t\right) + \gamma_t \cdot \delta (E,E^\star) 
\end{equation}
 with $g \left(E;t\right)$ being the DOS estimation on the $t$-th step, and $E^\star$ being the value of the sampling parameter at the end of step number $t$, i.e., $E^\star = E\left( x_1\right)$ if the $t$-th trial move was accepted, and $E^\star = E \left( x_0 \right)$ otherwise. The $\delta (E,E^\star)$ is the Kronecker delta, it takes the value $1$ only if both parameters are equal and $0$ otherwise. In the case of continuous parameters, their values should be replaced by the index of the corresponding bin. The value of the modification factor $\gamma_t$ depends on the trial moves counter $t$. In contrast to the original Wang-Landau algorithm \cite{wl1,wl2}, having problems with convergence \cite{TSconvergence,wl1t1,wl1t2}, the SAMC algorithm has mathematically proven convergence \cite{samc1,samc2}. The fastest functional convergence is reached when the modification factor depends on the step counter as
\begin{equation}
\label{eq:gamma_samc}
\gamma_t = \min\left[\gamma_0; \frac{t_0}{t} \right]
\end{equation}
 here $\gamma_0$ is pre-chosen maximal value of the modification factor, and $t_0$ is a parameter controlling the decay rate of the modification factor. An alternative definition of the modification factor is
\begin{equation}
\label{eq:gamma_samc2}
\gamma_t = \frac{t_0}{t_1 + t}
\end{equation}
with $t_1$ being a parameter defining initial value of the modification factor. The definitions (\ref{eq:gamma_samc}) and (\ref{eq:gamma_samc2}) have similar asymptotic behaviour at small and large counter values, as $\gamma_0= t_0 / t_1 $. In practice, typical values of $\gamma_0$ are $ \gamma_0 \le 10^{-1}$ or $t_1 \ge 10 \cdot t_0$. A good choice of $\gamma_0$ guarantees visiting (not necessarily uniformly) of a large (up to 100\%) part of the sampling parameter range within the initial stage of the simulation when $\gamma \approx \gamma_0$. At the same time, too-small values hinder compensation of possible inaccuracies of the initial DOS estimation, therefore choice of the $\gamma_0$ value requires a balancing of these opposite effects. In the rest of the paper, we use the relation (\ref{eq:gamma_samc2}).

\subsection{Models}
\label{sec:models}
We consider three models covering three possible combinations of discreteness/continuity of the configuration space and the energy spectrum. Discrete-discrete, continuous-discrete, and continuous-continuous pairs are represented correspondingly by the Ising model, the hard-sphere chain, and the system of Lennard-Jones particles.

\subsubsection{Ising model}
\label{seq:ising}
The system is located on the $2D$ square lattice with {  $N = 50^2$} sites. The energy of the system is given by
\begin{equation}
\label{eq:isingEn}
E = -\sum_{<ij>} s_i s_j
\end{equation}
with $<ij>$ denoting nearest neighbour pairs accounting for periodic boundary conditions, and $s_i$  describes a state of $i$-th site and takes values $0$ or $1$. The exact number of states for this model is calculated according to \cite{beale1996}.

\subsubsection{Hard-sphere chain}
\label{sec:hschain}

The linear chain consists of tangent hard-spheres with the fixed bond length $d$.
The energy of a chain conformation depends in this model only on the pair distances between the beads centers $r_{ij}$: $E = \sum_{i,j>i} u\left( r_{ij}\right)$ with the pair interaction energy $u\left( r_{ij}\right)$:
\begin{equation}
\label{eq:hs_en}
u\left( r_{ij}\right) = 
\begin{cases}
\begin{aligned}[t]

\infty,&& r_{ij} < d\\
-\varepsilon,&& d< r_{ij} < \lambda \cdot d\\
0, &&  \lambda \cdot d<{r_{ij}}  \\
\end{aligned}
\end{cases}
\end{equation}
here $\lambda>1$ defines the interaction range. The exact density of states for the $N=6$ beads is calculated in \cite{taylorExact} for a series of interaction ranges.

\subsubsection{Lennard-Jones particles}
\label{sec:ljpart}
The system of $N=13$ particles is placed in a cubic box with periodic boundary conditions. The energy of the system $E=\sum_{i,j>i} v\left(r_{ij} \right)$ is given by the non-truncated Lennard-Jones potential:
\begin{equation}
\label{eq:lj_en}
v\left( r_{ij}\right) = 
\varepsilon \left[ \left( \frac{\sigma}{r_{ij}}\right)^{12} - 2 \left( \frac{\sigma}{r_{ij}}\right)^{6}\right]
\end{equation}
with $\sigma$ being the distance at which the pair interaction reaches the minimal value $-\varepsilon$ and $r_{ij}$ being the minimal image pair distance between particles $i$ and $j$. The linear size of the simulation box is $L=30 \sigma$. We restrict the energy range for the $13$-beads LJ-system to $U< -\varepsilon$. In the present work we use two discretizations of the energy: with bin width $\Delta U = \varepsilon$ and $\Delta U = 0.1 \varepsilon$. For both discretizations the reference DOS for the convergence estimation was estimated during $4$ independent runs of length approximately $10^{12}$ trial moves each. The DOS, averaged over a set of independent runs, providse a good basis for the error estimation \cite{TSconvergence}.

\section{Results}

\subsection{Convergence properties of the Stochastic Approximation Monte Carlo algorithm}
\label{sec:samc_conv}

\begin{figure}[t]
{\includegraphics[width=0.95\columnwidth]{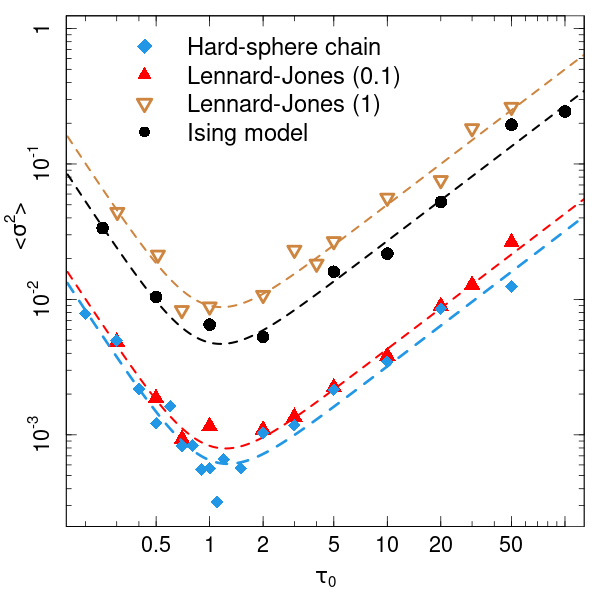}}
\caption{Final step mean-square errors averaged over 30 runs vs. modification factor decay rate, $\tau_0 = t_0 /N_\text{bin}$ for the Ising model (circles), Lennard-Jones system with the bin width $\Delta U=0.1 \varepsilon$ (filled triangles) and $\Delta U = \varepsilon$ (open triangles) and the hard-sphere chain with $\lambda=1.35$ (diamonds). Dashed lines show the fit of the data by $f=A \cdot \tau_0 + B \cdot \tau_0^{-2}$. Values of the fitting constants are summarized in the Supplemental materials. }
\label{fig:errorsSAMC}
\end{figure}

The SAMC algorithm with the modification factor decay rate (\ref{eq:gamma_samc}) or (\ref{eq:gamma_samc2}) converges to the exact DOS \cite{samc1,samc4,TSconvergence} with the mean square error proportional to the current value of the modification factor. We calculate the square-error of a SAMC run as
\begin{equation}
\label{eq:EstimationError}
\sigma^2 \left(t \right)=\frac{1}{N_\text{bin}} \sum_{n=1}^{N_\text{bin}} \left[ \ln g_{n} \left( t \right) - \ln \hat{g}_{n} \right]^2
\end{equation}
with $N_\text{bin}$ being the total number of discrete energy values or energy bins in the sampling range,
$g_{n} \left( t \right)\equiv g\left(E_n ; t\right)$ being the DOS estimation after $t$ trial moves,
and the $\hat{g}_n \equiv \hat{g}\left(E_n\right)$ being the reference DOS (exact for the Ising model and the hard-sphere chain, or estimated during an extremely long SAMC run for the system of Lennard-Jones particles). Both DOS are shifted to fulfill the condition $\sum_{n=1}^{N\text{bin}} \ln g_n =\sum_{n=1}^{N\text{bin}} \ln \hat{g}_n =0$ \cite{TSconvergence}. Thus with the decay rate (\ref{eq:gamma_samc}) or (\ref{eq:gamma_samc2}) the estimate of mean-square error decays as 
$\left\langle \sigma^2\left( t \right) \right\rangle = C/t \propto \gamma_t$ \cite{samc4,TSconvergence}, with the constant $C$ depending on the simulation parameters, e.g., decay rate $t_0$. Such behaviour allows us to consider  the error on the final step of a long run as a measure of the asymptotic convergence rate. 

The final accuracy of a flat-histogram simulation run depends on the total number of energy values (or energy bins) in the model system, $N_\text{bin}$, because all of them are expected to be visited approximately uniformly during the simulation. This makes the number of bins a natural scale for the Monte Carlo time: $\tau =t/N_\text{bin}$. Therefore Fig.~\ref{fig:errorsSAMC} represents the final mean-square errors as functions of the rescaled decay rates $\tau_0 = t_0 / N_\text{bin}$.
For all considered models, the final mean-square errors demonstrate non-monotonic behaviour with the minimum reached in the range $1< \tau_0 <2$.
Because of the stochastic nature of the algorithm, the error in each particular simulation is distributed around the mean value with variance comparable to the mean. Thus, on average, the accuracy of the results obtained with $\tau_0 =1.5$ or $\tau_0 =2.5$ will be comparable.  It is also significant to note, that  for both energy discretizations in the Lennard-Jones model the position of the minimum is determined by the number of bins, but not by the total energy range, which is the same for both discretizations. This confirms our choice of the number of bins as a scale for the Monte Carlo time.

\subsection{Flatness of the histogram as a relative convergence characteristic}
\label{sec:flatness}
\begin{figure}[!h]
{\includegraphics[width=0.95\columnwidth]{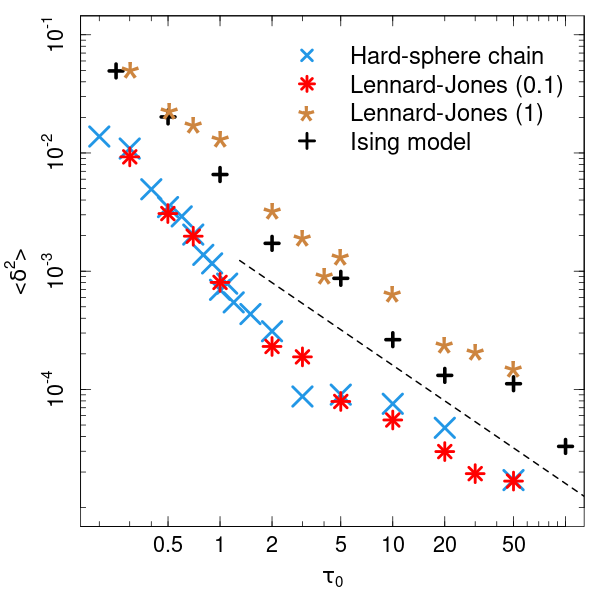}}
\caption{Final flatness of the visitation histogram averaged over 30 runs vs. modification factor decay rate, $\tau_0 = t_0 /N_\text{bin}$ for the Ising model (crosses), the Lennard-Jones system with the bin width $\Delta U=0.1 \varepsilon$ (5-arm stars) and $\Delta U = \varepsilon$ (8-arm stars) and the hard-sphere chain with $\lambda=1.35$ (diagonal crosses). The dashed line indicates $\propto 1/\tau_0$ decay.}
\label{fig:flatSAMC}
\end{figure}

"Flat-histogram algorithms" is a widely used name for this family of simulation techniques. The feature of approximately equal visiting (flatness of the visitation histogram) may be also a decision criterion in particular cases. For instance, in the original Wang-Landau algorithm \cite{wl1,wl2} the reaching of some lower bounds by the flatness of the visitation histogram governs the change of the modification factor value.
To investigate the role of the visitation histogram flatness for convergence estimation, we quantify it as the mean-square relative deviation of the histogram from its mean value:
\begin{equation}
\label{eq:HistogramFlatness}
\delta^2 \left(t \right)=\frac{1}{N_\text{bin}} \sum_{n=1}^{N_\text{bin}} \left[ \frac{H_{n} \left( t \right) - \langle H \rangle}{\langle H \rangle}\right]^2
\end{equation}
with $H_{n} \left( t \right)$ being the number of visits at the bin number $n$ during the $t$ trial moves since the start of the simulation, and with $\langle H \rangle = t/N_\text{bin}$ being the mean value of the visitation histogram after $t$ trial moves. Similar to the mean-square error of the DOS, the relative histogram deviation decreases as the simulation length grows \cite{wl1t1,samc_hist_flat}. But in contrast to the DOS error, for a given simulation length the averaged deviation from the histogram mean value decrease monotonically with $\tau_0$  (Fig.~\ref{fig:flatSAMC}). 

\begin{table}[!b]
\caption{Averaged plateau values of rescaled mean-square errors and histogram deviations from the mean for different models}
\begin{tabular}{ c c c c}
\hline\hline
{Model} & ~{$10^3 \cdot \langle \sigma^2 \rangle / \tau_0$}~ & ~{$10^3 \cdot \langle \delta^2 \rangle \cdot \tau_0$}~ \\ \hline
Ising model & $2.8 \pm 0.6$& $3.7 \pm 1.3$ \\ 
Lennard-Jones (1) & $5.5 \pm 1.2 $ & $5.9 \pm 1.2 $\\ 
Lennard-Jones (0.1) & $0.45 \pm 0.05$ & $0.59 \pm 0.14$  \\ 
~Hard-sphere chain ($\lambda =1.35$)~ & $0.37 \pm 0.08$  & $0.65 \pm 0.29$ \\ \hline
\hline
\end{tabular}
\label{tab:SAMCerrorPlat}
\end{table}

The opposite behaviour of visitation histogram deviations and the DOS  error impedes using the histogram flatness as an absolute quantitative measure of algorithm convergence. Nevertheless, mean values of these parameters remain related as they originate in the same sampling process. To eliminate the opposite effect of $\tau_0$, we rescale both functions by dividing and multiplying by $\tau_0$ of the DOS error and histogram deviations correspondingly. Table~\ref{tab:SAMCerrorPlat} summarizes the plateau values of the rescaled functions for $\tau_0 > 1$.
Similar within one standard deviation plateu values allow for a coarse estimation of the mean error of a series of SAMC runs having the same parameters: $\langle \sigma^2 \rangle \approx \langle \delta^2 \rangle \cdot \tau_0^2$.
However, for each particular run, the variation of the histogram can vary over one order of magnitude for the same DOS error (see Supplemental Materials).
Therefore, the flatness of the SAMC visitation histogram does not provide relevant information about the accuracy of a particular single run: $ \sigma^2 \left(t\right) \neq \delta^2 \left(t\right) \cdot \tau_0^2$.

\subsection{Enhanced convergence}
\label{sec:ImportanceSampling}

\begin{figure}[!t]
{\includegraphics[width=0.95\columnwidth]{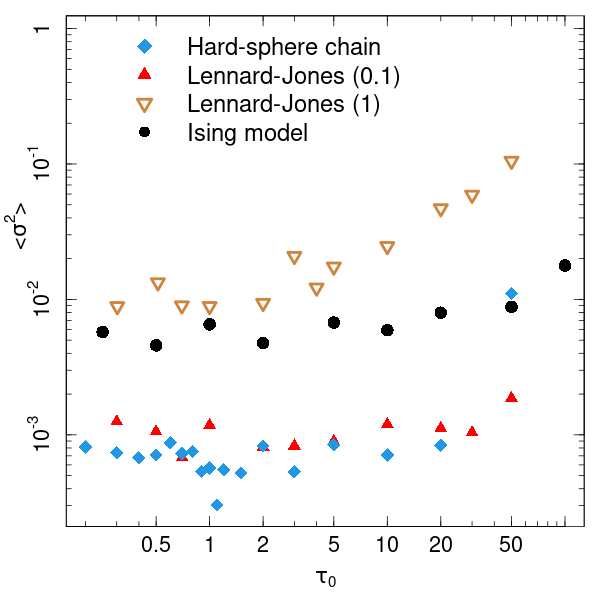}}
\caption{Final mean-square errors, $\langle \sigma^2 \rangle$, for the importans sampling procedure averaged over 30 runs vs. modification factor decay rate, $\tau_0 = t_0 /N_\text{bin}$. The symbol notations are the same as in the Fig. \ref{fig:errorsSAMC}.}
\label{fig:impSampl}
\end{figure}

\begin{figure}[!t]
{\includegraphics[width=0.9\columnwidth]{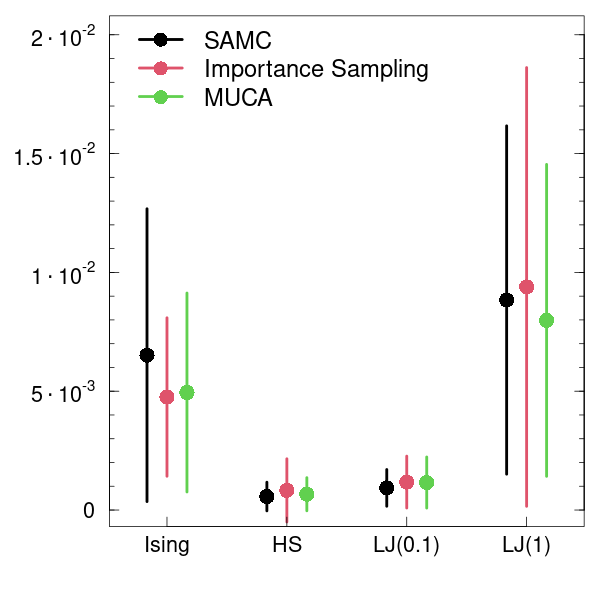}}
\caption{Mean-square errors calculated for the Ising model, the hard-sphere model and the Lennard-Jones model with the bin width $\Delta U =0.1 \varepsilon$ (LJ0.1) and $\Delta U = \varepsilon$ (LJ1) for the SAMC simulation with close to the optimal parameter choice $\tau_0=1$ (black), for the importance sampling procedure with another underlying SAMC parameter $\tau_0=2$ (red) and for the MUCA sampling (green). The points indicate the mean-square error, and vertical lines correspond to the standard deviation. The results were calculated for 30 independent runs.
}
\label{fig:errorCompare}
\end{figure}

Flat-histogram algorithms utilize a pre-estimated or, during the run, modified non-normalized probability or density of states to sample a broad range of parameter values. From that point of view, these algorithms could be described in terms of importance sampling. The flat-histogram algorithms replace the uniform distribution of micro-states or points in conformational space, $p\left(x \right)=\tilde{p}={const}$, by the approximately uniform distribution with respect to the sampling parameter, e.g. energy.
This leads to the new probability distribution in configurational space $q\left(x \right) \propto 1/g\left( E\left( x \right)\right)$. Then an integration with weights $p\left(x\right)$ of a function $O\left(x\right)$ could be replaced by an estimation of the modified function calculated with weights $q\left( x \right)$:
\begin{equation}
\label{eq:ImportanceSamplingDefinition}
\left\langle O\left( x\right) \right\rangle_{p\left( x \right)} = \left\langle O\left( x\right) \frac{p\left(x\right)}{q\left(x\right)} \right\rangle_{q\left( x \right)} 
\end{equation}
here $\left\langle \dots \right\rangle_{f\left( x \right)}$ denotes the averaging with the distribution function $f\left( x\right)$.

On the other hand, the probability to observe a parameter value from the given bin $n$ is
\begin{equation}
\label{eq:Importance0}
g^{\text{IS}}_n = \int dx~I_n\left(x\right) \cdot p\left(x\right) = \left\langle I_n\left(x\right)\right\rangle_{p\left( x \right)} 
\end{equation}
here $I_n\left(x\right)$ is the indicator function, which takes the value $1$ if $x$ parameter value corresponds to the bin number $n$, and takes value $0$ otherwise. Combining relations (\ref{eq:ImportanceSamplingDefinition}) and (\ref{eq:Importance0}), we get the relation for the uniform distribution in configuration space
\begin{equation}
\label{eq:Importance1}
\ln g^{\text{IS}}_n = \ln \tilde{p} + \ln \left\langle \frac{I_n\left(x\right)}{q\left(x\right)}\right\rangle_{q\left( x \right)} 
\end{equation}
Practically, we can only estimate this mean value during the flat-histogram run. Thus taking into account the definition of $q(x)$ the practical relation is
\begin{equation}
\label{eq:Importance2}
\ln g^{\text{IS}}_n \approx const + \ln \sum_{t} {\left[I_n\left(x_t \right) \cdot {g \left(E_n;t\right)} \right]} 
\end{equation}
where the summation covers all trial moves since the start of the simulation, $x_t$ is the configuration on the end of trial move $t$,
and the constant $const$ includes logarithms of normalization constants.

Implementation of the importance sampling estimation (\ref{eq:Importance2}) provides minimal computational costs. It requires only accumulation of $\ln g^{\text{IS}}_n$ before the SAMC DOS increment. To avoid an overimpact of the final values of the  growing DOS, we increment the importance sampling estimation with the SAMC DOS normalized similar to the error calculation: $\sum_{n=1}^{N\text{bin}} \ln g_n =0$. Fig.~\ref{fig:impSampl} illustrates accuracy reached with the importance sampling accumulation. Typically, the importance sampling provides similar or better accuracy than the SAMC procedure underlying this estimation.
Moreover in most cases the importance sampling accuracy depends not on the error of the underlying SAMC or the $\tau_0$ value. Only in the case of the Lennard-Jones model with broad bins ($\Delta U = \varepsilon$) or largest $\tau_0$ does the importance sampling error grow significantly. The cause of this requires a deeper investigation, but we assume that stronger fluctuations of the underlying  SAMC DOS estimation generate large numerical noise, which could disturb the estimation on the initial stage of simulation and requires longer runs to be compensated. The numerical origin of these difficulties could be related to large modification factor values and bin width which slowdown the energy change and induce longer accumulation of the values within single bin. We suppose that in both cases, smaller initial modification factor values could decrease the impact of these disturbing factors. 

In the case of MUCA the summation in eq.~(\ref{eq:Importance2}) is trivial. Since all summands for a given $n$ are equal during the complete run, the sum equals to the underlying DOS multiplied by the visitation histogram:
\begin{equation*}
\ln g^{\text{IS}}_n \approx const + \ln  {g \left(E_n\right)} + \ln  {H \left(E_n\right)} 
\end{equation*}
which is equivalent to the conventional multicanonical DOS estimation (\ref{eq:muca}).

This allows a comparison of the accuracy of three considered DOS estimation procedures: the SAMC simulation with an approximately optimal choice of $\tau_0$, the importance sampling based on an underlying SAMC simulation, and the MUCA simulation. Errors of all three approaches are similar within one standard deviation, whose typical value is of the same order as the averaged error (Fig.~\ref{fig:errorCompare}). Consiquently, if the optimal parameters are selected, no significant improvement of the accuracy can be reached by the choice of the simulation approach. On the other hand, with SAMC sampling utilizing a non-optimal decay rate, importance sampling accumulation allows greater flexibility in SAMC-parameter choice.

\subsection{Extended sampling}
\label{sec:Extensions}
The combination of the flat-histogram approach with importance sampling has a broader application than merely weakening the influence of parameters choice on the error of the estimated DOS. The importance sampling accumulation is not reduced to exactly the same bins as in the underlying flat-histogram sampling. In this section, we consider a few examples of alternative bin choices which permit extracting more information during one run or improving the accuracy of algotithms through modification of the function determining the parameter values.

The energy of a system is a function characterized by a set of implicit and explicit parameters, for instance, interaction range or typical energy scale. A change of parameters may change the energy for a given configuration. On the other hand, this opens the possibility to extract information concerning changes in the model system associated with the energy parameters change without new simulations. A new bin choice requires minor modification of the importance sampling accumulation (\ref{eq:Importance2}). New weights $w_n\left( c \right)$ could be estimated as:
\begin{equation}
\label{eq:ImportanceNEW}
\ln w_n \approx const + \ln \sum_{t} {\left[J_n\left(x_t \right) \cdot {g_t \left(E_n\right)} \right]} 
\end{equation}
with $J_n\left(x_t \right)$ being the indicator function describing the new bins. Because of the difference in the bins between the underlying flat-histogram simulation and the importance sampling ones, the sum in eq.~(\ref{eq:ImportanceNEW}) may include non-equal terms even with an unchanged DOS during a MUCA run and cannot be replaced in this case by the visitation histogram.

\subsubsection{Altering the interaction range}
\label{sec:HS132}
This section  exemplifies the importance sampling accumulation for the interaction range parameter values. We denote the weight function for the interaction range parameter $\lambda$ as $w_{\lambda}\left(E_n\right)$ with $E_n = E_n \left( \lambda\right)=-\varepsilon n$, i.e. each bin corresponds to a single energy value calculated for the given $\lambda$ according to (\ref{eq:hs_en}). We perform all flat-histogram simulations with $\lambda=1.35$. Thus, estimation of $w_\lambda$ according to (\ref{eq:ImportanceNEW}) requires an update only in one bin after acceptance or rejection of the trial move $t$:
\begin{equation}
\label{eq:ImportanceHS}
w_\lambda \left( E  \left( x_t ; \lambda \right)  \right)  \rightarrow
w_\lambda \left( E \left( x_t ;\lambda \right)  \right) + g_{t} \left(E \left( x_t ; 1.35\right)\right)
\end{equation}
here $E \left( x_t ;\lambda \right)$ denotes the energy calculated for the conformation of the chain at the end of the trial move $t$ according to (\ref{eq:hs_en}) with interaction range $\lambda$. It is necessary to underline that energies calculated for different interaction ranges may differ, thus bin indexes determined for $w_\lambda$ and $g_t$ do not necessarily coincide, and that the update (\ref{eq:ImportanceHS}) precedes the update of $\ln g\left(x_t \right)$.

\begin{figure}[!t]
{\includegraphics[width=0.9\columnwidth]{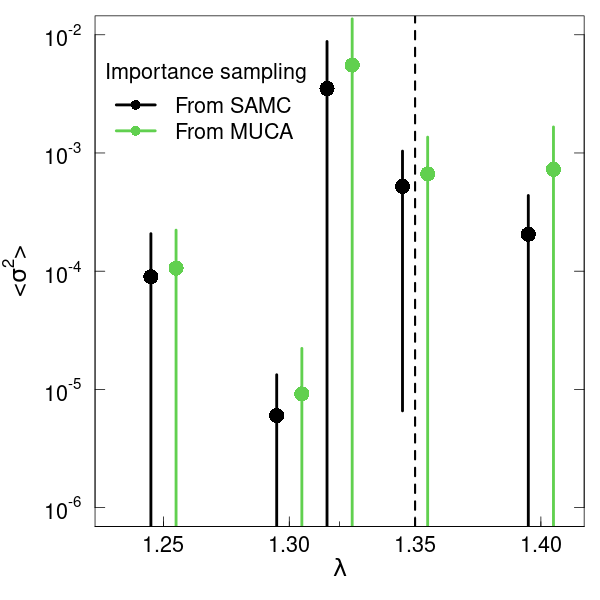}}
\caption{Mean-square errors for importance sampling estimation with underlying MUCA and SAMC flat-histogram samplings calculated for the hard-sphere chain with different interaction ranges. The results are averaged over 30 runs with underlying sampling having interaction range $\lambda=1.35$ (indicated by vertical dashed line).
}
\label{fig:errorLambda}
\end{figure}

Importance sampling based on underlying flat-histogram simulation (SAMC or MUCA) may provide an estimation of DOS for any number of interaction ranges within one run with the fixed interaction range. Fig.~\ref{fig:errorLambda} shows calculated errors for a series of interaction ranges of the 6-mer hard-sphere chain reported in \cite{taylorExact}. For all presented interaction ranges, the errors of the DOS estimated for both types of underlying flat-histogram simulations (SAMC and MUCA) are similar within the one standard deviation. Only for single interaction range presented in Fig.~\ref{fig:errorLambda} is the error of the importance sampling estimation significantly larger than the error of the underlying process: $\lambda=1.32$. 
This interaction range is located close to the boundary separating different minimal energies in the system: $-8 \varepsilon$  and $-9 \varepsilon$ \cite{taylorExact}. Since the interaction range $\lambda=1.32$ is extremely close to the change point, the number of states corresponding to the first excited state ($-8 \varepsilon$) is $\approx 2.57\cdot 10^7$ times larger than in the ground state ($-9 \varepsilon$). This makes direct sampling of this interaction range more complicated than for the next presented interaction range $\lambda=1.35$, where this ratio is only $\approx 7.79\cdot 10^3$. Despite the large difference in DOS values, the importance sampling with the underlying flat-histogram simulation for $\lambda_0 =1.35$ provides a good estimation of the DOS for $\lambda_1=1.32$. In contrast, for the direct sampling of the same length ($10^{10}$ trial moves) we cannot provide an error estimation because the lowest energy state was found only in 3 simulations of 30 for the $t_0 =15$ ($\tau_0 =1.5$) and was not found in 30 runs with $t_0=10$ ($\tau_0 =1.0$ ).

\subsubsection{Altering the relative energy contributions}
\label{sec:EnergyRescale}

A further example of energy parameters is the relative contribution of different energy terms in the total configurational energy. In this section we consider the energy of the Lennard-Jones system in a box with periodic boundary conditions along $x$- and $y$-axes and impenetrable walls in the $z$-direction. One of the walls is reflective and excludes translations, but does not contribute to the energy. Interaction with the second wall (located at $z=0$) is described by a 1-dimensional 12-6 Lennard-Jones potential:
\begin{equation}
\label{eq:wall}
E_w = 
\varepsilon \sum_k  \left[ \left( \frac{\sigma}{z_{k}}\right)^{12} - 2 \left( \frac{\sigma}{z_{k}}\right)^{6}\right]
\end{equation}
here the sum includes all particles, $z_k$ is the $z$-coordinate of the $k$-th particle and the parameters $\varepsilon$ and $\sigma$ are the same as in the definition (\ref{eq:lj_en}).

The conformational energy of the system is the sum of two contributions:
\begin{equation}
\label{eq:sumEn}
E_{\alpha} = E_{\text{LJ}} + \alpha \cdot E_w
\end{equation}
with $E_{\text{LJ}}$ being sum of pair interactions (\ref{eq:lj_en}), and the constant $\alpha$ describing strength of the interaction with the attractive wall.
We accumulate the importance sampling estimation for the altered attraction strength as follows:
\begin{equation}
\label{eq:ISattrWall}
w\left( E_\alpha\right)  \rightarrow w \left( E_\alpha \right)  + g_{t} \left(E_{\text{LJ}} + \alpha_0 E_w \right)
\end{equation}
with $w \left( {E_\alpha} \right) = w\left( E_{\text{LJ}} + \alpha E_w \right)$ being the importance sampling estimation of the DOS for the wall interaction strength $\alpha$, and $\alpha_0$ being the wall interaction strength in the underlying flat-histogram simulation.

\begin{figure}[!t]
{\includegraphics[width=0.9\columnwidth]{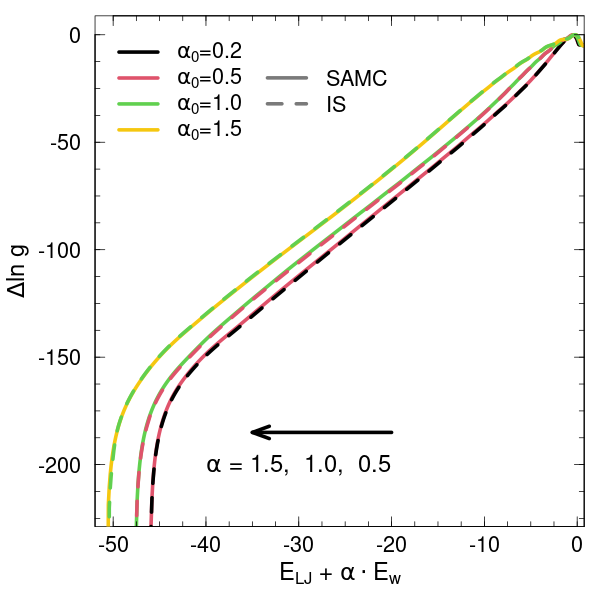}}
{\includegraphics[width=0.9\columnwidth]{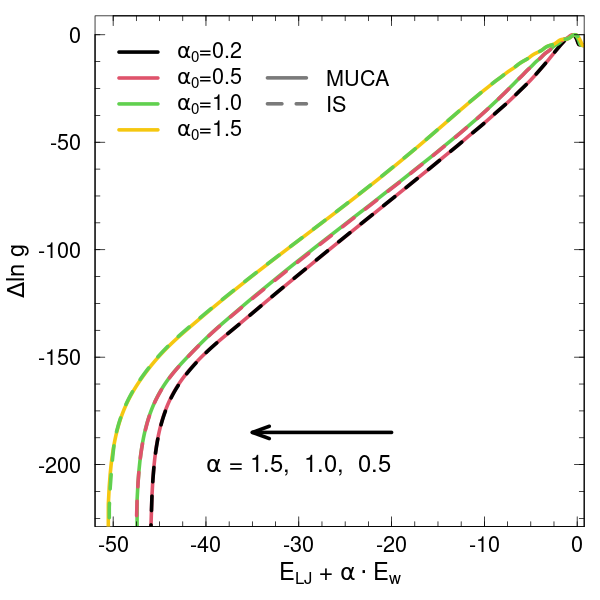}}
\caption{Comparison of importance sampling (dashed lines) and the underlying flat-histogram sampling: SAMC (top) and MUCA (bottom), for the Lennard-Jones system in a box with attractive wall with different particle-wall interaction strengths (for lines from the right to left: $\alpha =0.5,~ 1.0,~ 1.5$). The color of the curves corresponds to the interaction strength of the underlying flat-histogram simulation, $\alpha_0$. The logarithms of DOSs are shifted with respect to their maximum: $\Delta \ln g = \ln g -\max\left[ \ln g \right]$. Results are averaged over 10 runs.
}
\label{fig:wallAttraction}
\end{figure}

Estimations based on importance sampling with altered interaction and based on either flat-histogram algorithms are in a good agreement (see Fig.~\ref{fig:wallAttraction}). It is natural, that with the increase of the difference between altered and underlying interaction strength, the accuracy of DOS estimation decreases. Therefore, we compare in Fig.~\ref{fig:wallAttraction} only results observed for the nearest pairs of $\alpha$. Despite the fact that for $\alpha$ stronger deviating from $\alpha_0$, the accumulated IS DOS estimations are less accurate, they can be utilized as the initial estimation of the DOS in a flat-histogram simulation. Importance sampling accumulation provides also an upper bound estimation of the ground-state energy for the considered $\alpha$.

\subsubsection{Pressure estimation}
\label{sec:Pressure}
The simulation box size determines not only the system volume, $V$, but also affects the energy through periodic boundary conditions. From the statistical physics point of view the connection between these parameters and the microcanonical entropy, $S\left( E, V\right) = k_{B} \ln g\left(E, V\right)$, determines the pressure in the system:
\begin{equation}
\label{eq:pressure2}
\beta p = \left( \frac{\partial \ln g}{\partial V} \right)_E
\end{equation}
here $p$ is the pressure, and $\beta = (k_B T)^{-1} = \partial \ln g / \partial E$.

The volume dependence of the DOS can be estimated by altering the volume of the simulation box with importance sampling, taking into account boundary conditions. For a simulation box with periodic boundary conditions, a given particle configuration could be described by an infinite number of coordinates differing by box-size translations. To exclude this degeneracy, we consider only those coordinates translated to the box as the unique main set of coordinates describing the given particle configuration.
With this restriction, if a given set of coordinates is the main one for the given box size, then numerically the same coordinates also describe a main set in any larger box. This is not necessarily true for smaller boxes. As we consider a smaller box where one of particles is located outside of it, translation is required to bring this particle inside the box of smaller size.
But the translated coordinates set has another corresponding main set in the initial box.  This allows for estimation of the volume (or box size) assocciated contributions to the DOS. 

As a simple practical realization of the volume sampling approach, we accumulate the importance sampling estimate for the volume, changed by reducing the simulation box size along the $z$-axis: $L_z=L-\Delta L$.
 As the main set we consider positive $z$-coordinates of particles not exceeding the box size $0 < z_k \le L_z$, then the importance sampling accumulation takes the form:
\begin{equation}
\label{eq:ISvolume}
w\left( E^\prime;V \right)  \rightarrow w \left( E^\prime; V\right)  + g_{t} \left(E;V_0\right) \cdot \Theta \left( L_z- z_{\text{max}} \right)
\end{equation}
here the importamce sampling box volume $V=L^2 \cdot L_z$, and the underlying flat-histogram simulation box volume $V_0 = L^3$, $E^\prime$ and $E$ are corresponding energies,  and  $\Theta\left( L_z- z_{\text{max}} \right)$ with $z_\text{max} = \max\{ z_k\} $ is the Heaviside step function. The step function is responsible for the exclusion of coordinates sets, which are not main ones within the smaller box, i.e. they have at least one particle outside the smaller box.
Generally, the energy depends on the box size because of periodic boundary conditions and should be calculated for each box size. But if volume variation is small we can neglect this effect and take $E^\prime \approx E$ in the sence that both energies belong to one the same energy bin.

\begin{figure}[!t]
{\includegraphics[width=0.9\columnwidth]{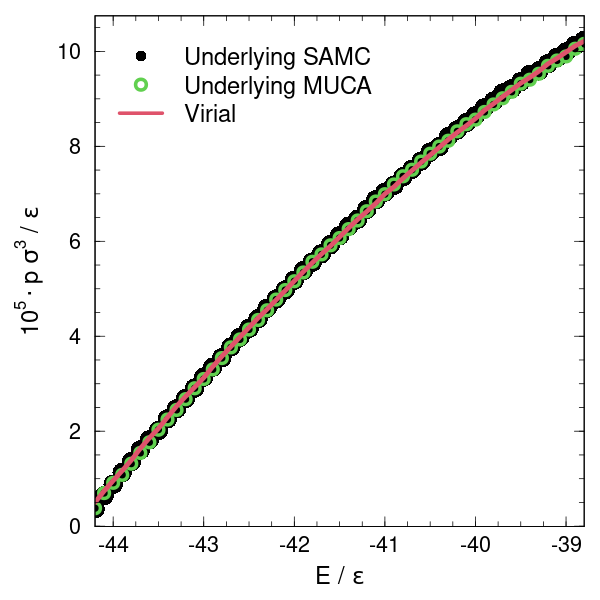}}
\caption{Estimation of pressure calculated through importance sampling accumulation as a function of volume for the underling SAMC (filled circles) and the MUCA (empty circles) simulation, and the pressure calculated from the virial (line) for $13$ Lennard-Jones particles in a cubic box of linear size $30\sigma$ with periodic boundary conditions. The results are averaged over 10 independent runs.
}
\label{fig:pressure}
\end{figure}

As an example we consider the Lennard-Jones system in the cubic simulation box with periodic boundary conditions and the linear size $L=30\sigma$. To estimate the volume dependence of the DOS we accumulate the importance sampling estimations (\ref{eq:ISvolume}) for a set of volumes $V_m=L^2 \cdot (L - \Delta L_{zm})$ with $\Delta L_{zm} =m \cdot \delta L_z$, $\delta L_z=10^{-2}\sigma$ and $m=0,1\dots 9$. The volume dependence was fitted by a linear function independently for each energy bin, so the slope of the fit estimates the volume derivative (see Supplementary materials). Since the derivative of the DOS (\ref{eq:pressure2}) determines the product of the pressure and the inverse temperature we divide the estimated slope by the inverse temperature $\langle \beta \left(E\right)\rangle$ averaged over 10 runs. An independent estimation of the averaged pressure within each energy bin can be obtained from the virial:
\begin{equation}
pV_0 = - \frac{1}{3} \sum_{ij} \frac{\partial v\left( r_{ij}\right) }{\partial{r_{ij}}} r_{ij}
\label{eq:virial}
\end{equation}
with the summation over all pairs of particles $i$ and $j$, the pair separation $r_{ij}$ and the pair interaction energy $v\left( r_{ij}\right)$ calculated according to (\ref{eq:lj_en}). In Fig.~\ref{fig:pressure}, we include only pressure estimations for low energies, which correspond to the low-temperature single phase region of the system's phase diagram. For both types of underlying flat-histogram algorithms the pressure estimation is in excellent agreement with the virial calculations. The good agreement of the virial and volume derivative based results opens a way of relative simple pressure estimation in isochoric Monte Carlo sampling. This approach does not require the force calcuation and could be easily applied in models with fixed bonds or other features complicating the calculation of forces.


\section{Discussion and conclusion}

The flat-histogram algorithms are Monte Carlo algorithms sampling a broad energy range within a single simulation by utilizing an estimation of the density of states (DOS). These algorithms have been developed since half a century, and are known in many particular realizations. We group the variety of the suggested algorithms in two types. Algorithms of the first type require a frequent (after each trial move) update of the DOS, whereas algorithms of the second type hold the DOS unchanged during the simulation and re-estimate the DOS using the information from the complete run. We label these algorithms correspondingly as WL-like and as MUCA-like. 

We have compared the convergence of algorithms of both types and suggest possible extensions of their applicability. For a reasonable comparison we analyzed the optimal parameter choice for the stochastic approximation Monte Carlo algorithm (SAMC), which exemplifies the convergeing WL-type algorithms. We found that the optimal rate of the modification factor decay is determined by the number of energy bins, but not by the width of the investigated energy window (Fig.~\ref{fig:errorsSAMC}). Moreover the accuracy (mean-square error) of an optimal flat-histogram simulation depends not on the type of algorithm: the MUCA and the optimal SAMC simulations demonstrate similar errors for all investigated models. 

The flatness of the visitation histogram is considered as a key property of these algorithms, on which their naming is based. Despite this the flatness within a finite run is unreachable and weakly correlated with estimation accuracy. For the MUCA simulation, the flatness of the visitation histogram is governed by the accuracy of the initial DOS. Therefore an ideal flatness for an infinitely long simulation could be reached only with the exact DOS, which is unknown in practically interesting cases. In a SAMC simulation of given length, the histogram tends to become flatter as $t_0$ increases (Fig.~\ref{fig:flatSAMC}). This means that the averaged histogram flatness inversely correlates with the DOS accuracy. But owing to the exactly opposite behavior the averaged histogram flatness may provide a coarse estimation of the DOS error: $\langle \sigma^2 \rangle \approx \langle \delta^2 \rangle \cdot \tau_0^2$. It is necessary to underline, that this relation is applicable only for averaged error and histogram flatness, but not for parameters observed during each single run.

Flat-histogram algorithms are related to the importance sampling approach and could be utilized as a source of modified distribution in configuration space (eq. (\ref{eq:Importance2})). For the MUCA simulations, the importance sampling (IS) is just an alternative formulation, which provides exactly the same estimation of the DOS. For the SAMC approach, the IS accumulation provides a different and typically more accurate estimation of the DOS as compared to the underlying SAMC. Formally, in the limit $t_0 \rightarrow 0$ the SAMC-based importance sampling converts to the MUCA approach. Thus, the IS with underlying SAMC provides an estimation of the DOS that is independent of SAMC parameters at least in the range $\tau_0 < 5$. Moreover the IS accumulated estimation accuracy is similar to the MUCA and optimally converging SAMC ones. For large $\tau_0$ we expect that the IS estimation will depend on the initial value and the decay rate of the modification factor, but this problem requires a deeper investigation and is not addressed in this paper.

In many cases, investigation of a physical system is not restricted to a single set of parameters characterizing the system or interactions between particles. The broader investigation typically requires a series of runs with different parameters or an estimation of a multidimensional DOS. Both approaches take long computational times for independent runs or to reach the convergence of the DOS estimation. 
This problem could be partly solved by combining the flat-histogram algorithms with the importance sampling approach. The bins selection of the importance sampling accumulation is independent of the flat-histogram one and allows for the altering of the energy definition through a change of parameters of the potential. The change of parameters could concern the relative contributions of different energy terms (Sec.~\ref{sec:EnergyRescale}) or the spatial scale of the potential (Sec.~\ref{sec:HS132}). In some cases, the underlying sampling with another set of parameters could improve the convergence of the estimation by a more concentrated sampling of a part of configuration space that is difficult to investigate (see discussion in the Sec.~\ref{sec:HS132}).

Combination of the importance sampling and the flat-histogram approach could also be used for the estimation of derivatives of the DOS with respect to the system parameters. For instance, the DOS derivative with respect to volume (Sec.~\ref{sec:Pressure}) determines the pressure in the microcanonical ensemble. This definition does not require force calculations, which could be a complicated numerical problem in some models, e.g., for systems with fixed bond length.
In the case of dilute systems or small volume variations the estimation of the volume effects on the DOS could be simplified. Because of small volume variations, we suppose that energy change due to volume reduction is negligible and does not change the energy bin index associated with the configuration. In this case, the difference in the DOS contributions  originates from the exclusion of configurations having particles too close to the box walls, which, because of this, are not allowed for smaller boxes. Numerically, it is similar to the estimation of the size of an empty layer close to the box walls. A similar approach for the lattice models was suggested in \cite{latticePressure} and for a continuous polymer model in \cite{TSaggregation}. The possibilities of IS accumulation extension are not limited to the discussed examples and could be applied for the estimation of the chemical potential, the role of stiffness in polymer models, etc.

\begin{acknowledgments}
The author acknowledges funding by the German Science Foundation (DFG) under project number 189853844 (SFB-TRR 102). The author acknowledges Wolfgang Paul for stimulating discussions. Last but not least, I acknowledge my daughter for the time during her kindergarten adaptation, when I read a large volume of literature, none of which was applied in the present investigation directly, but stimulated the investigation.
\end{acknowledgments}

\include{bibl}

\newpage

\title{Flat-histogram algorithms: optimal parameters and extended application.\\~\\Supplemental Materials.}

\author{Timur Shakirov}
\affiliation{Institute of Physics, Martin Luther University of Halle-Wittenberg, Halle, Germany}

\maketitle

\renewcommand{\theequation}{S\arabic{equation}}
\renewcommand{\thefigure}{S-\arabic{figure}}
\renewcommand{\thetable}{S-\Roman{table}}
\renewcommand{\bibnumfmt}[1]{[S#1]}
\renewcommand{\citenumfont}[1]{S#1}

\onecolumngrid

\section{Fitting of the SAMC error dependency on the decay rate of the modification factor}
\begin{table}[!h]
\caption{Fiting of final mean-square errors for different models (see Fig.~1 in the main text) by the  function {$f= A \cdot \tau_0 + B\cdot \tau_0^{-2}$}.}
\centering
{$f= A \cdot \tau_0 + B\cdot \tau_0^{-2}$}\\
\begin{tabular}{ c c c c}
\hline\hline
{Model} & ~~{$10^3 \cdot A $}~~ & ~~{$10^3 \cdot B$}~~ \\ \hline
Ising model & ~~$2.65 \pm 0.19$~~& ~~$2.12 \pm 0.56$~~ \\ 
Lennard-Jones (1) & $5.09 \pm 0.29 $ & $3.82 \pm 0.68 $\\ 
Lennard-Jones (0.1) & $0.44 \pm 0.03$ & $0.39 \pm 0.04$  \\ 
~Hard-sphere chain ($\lambda =1.35$)~ & $0.30 \pm 0.02$  & $ 0.33 \pm 0.033$ \\ \hline
\hline
\end{tabular}
\label{tab:SAMCerrorPlat}
\end{table}

\vspace{-0.5cm}
\section{Volume dependency of the density of states}
\begin{figure}[h]
\begin{minipage}[c]{1\columnwidth}
{\includegraphics[width=0.8\columnwidth]{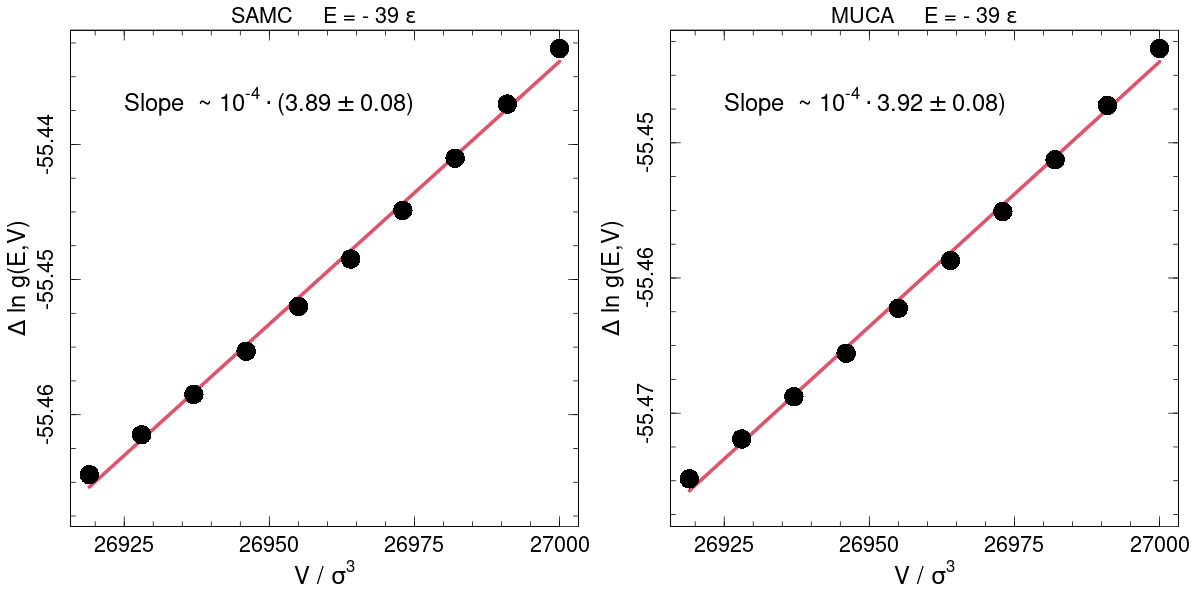}}
\end{minipage}
\begin{minipage}[c]{0.95\columnwidth}
\caption{DOS estimation for different volumes estimated with SAMC (left) and MUCA (right) algorithms. The energy bin is the same for both algorithms and corresponds to the energy range $-39\varepsilon \le E < -38.9 \varepsilon$. Red lines represent linear fit of the data. Slope of the fitting lines is given in figures. The inverse temperature is estimated as finite difference of $\ln g$ for two adjacent bins is $\beta\varepsilon= 3.86 \pm 0.08$ and $\beta\varepsilon= 3.86 \pm 0.09$ for SAMC and MUCA, respectively. This corresponds to the pressure estimation $p \approx \sigma^3 /\varepsilon \left( 1.01 \pm 0.03 \right) \cdot 10^{-4}$ and $p \sigma^3 /\varepsilon \approx \left( 1.02 \pm 0.03 \right) \cdot 10^{-4}$ for SAMC and MUCA, respectively.}
\end{minipage}
\end{figure}
The linear fit is equivalent to the Tailor expansion approximation restricted to the first derivative. Thus the slope of the fitting line provides an estimation of the derivative and allows to estimate the microcanonical pressure.

\section{Histogram flatness and DOS estimation accuracy}

\begin{figure}[t]
\begin{minipage}[c]{1\columnwidth}
{\includegraphics[width=0.65\columnwidth]{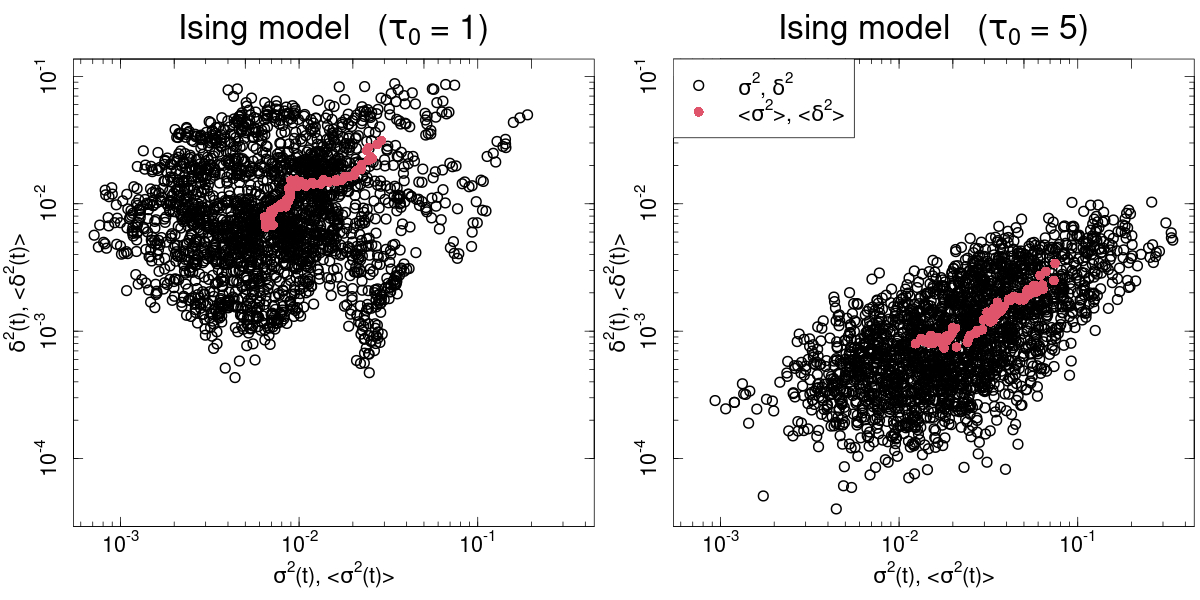}}
{\includegraphics[width=0.65\columnwidth]{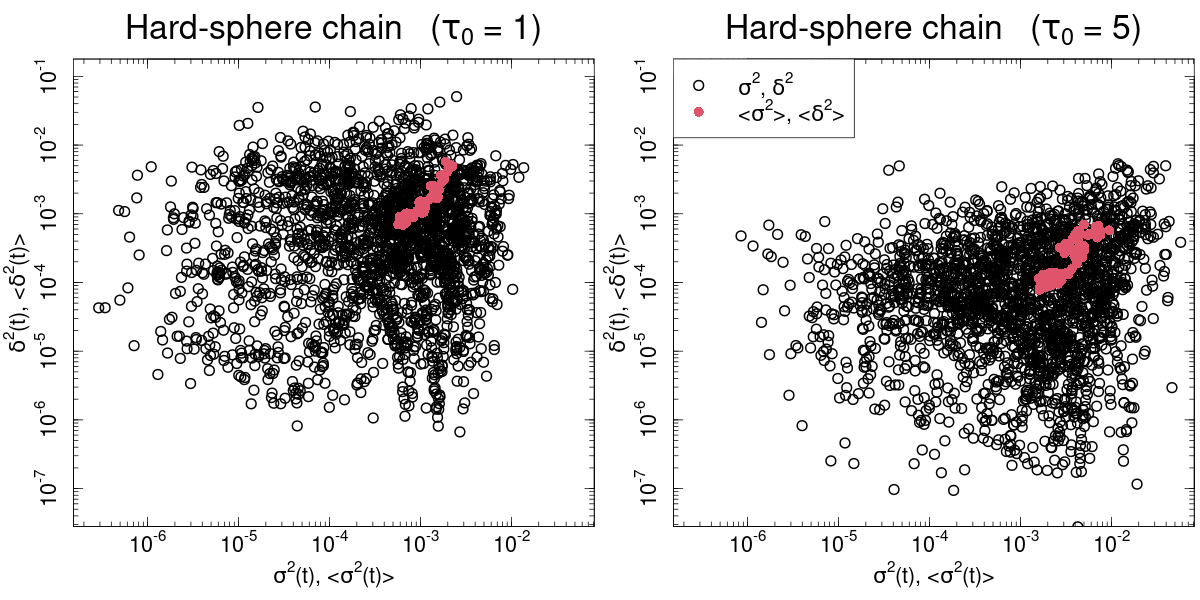}}
{\includegraphics[width=0.65\columnwidth]{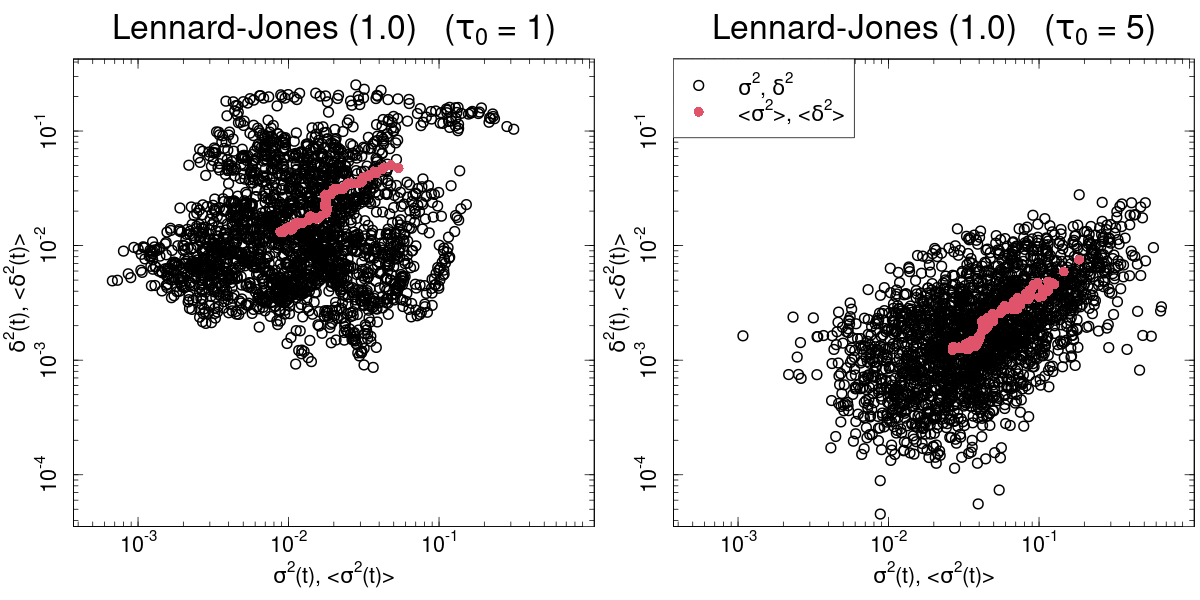}}
{\includegraphics[width=0.65\columnwidth]{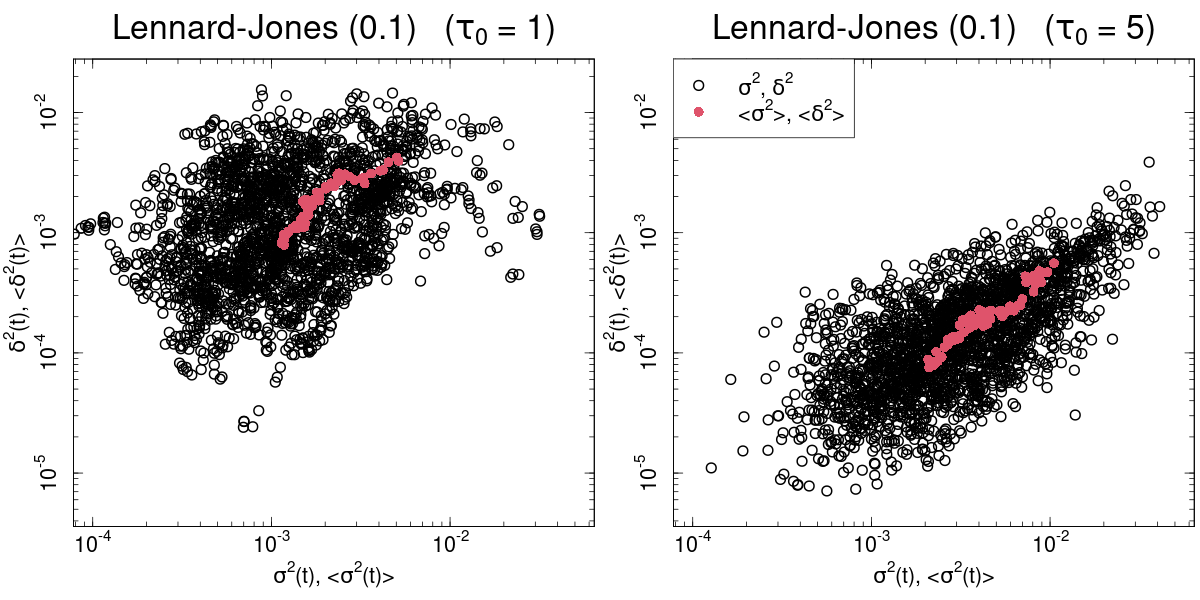}}
\end{minipage}
\begin{minipage}[c]{1.05\columnwidth}
\caption{Squares of histogram flatness ($\delta^2$) and DOS errors ($\sigma^2$) for different models and $\tau_0$ values. Open symbols represent the current pair of values $\left(\sigma^2,\delta^2 \right)$ for the last 80\% of simulation time for each of the 30 independent runs for each model and $\tau_0$. The values are calculated every $10^{-2} T$ steps, with $T$ being the total number of trial moves in a simulation. The filled symbols represent the averaged values of the functions $\left( \langle \sigma^2 \rangle,\langle \delta^2 \rangle \right)$. The averaging is performed over all 30 simulations, within the same time steps as those for the open symbols.}
\end{minipage}
\end{figure}

\begin{figure}[h]
{\includegraphics[width=0.66\columnwidth]{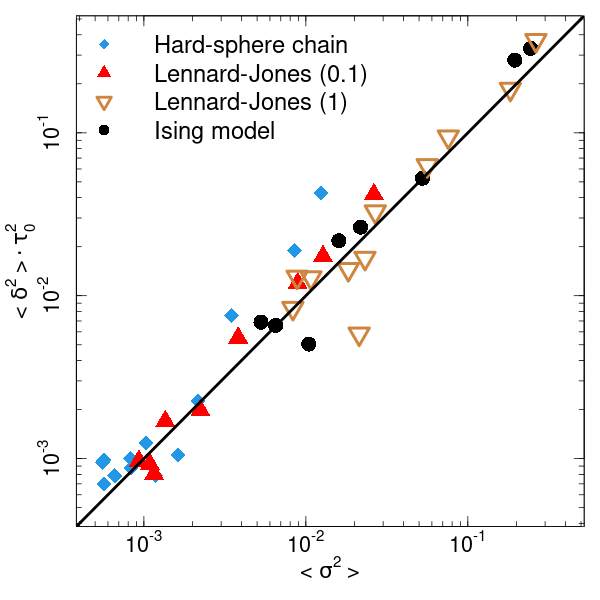}}
\begin{minipage}[c]{0.95\columnwidth}
\caption{Rescaled final histogram flatness parameter, $\langle \delta^2 \rangle \cdot \tau_0^2$, vs. final mean-square error of corresponding SAMC runs, $\langle \sigma^2\rangle$. The models are marked similar to the main text, and the results are averaged over $30$ independent runs for each point. The solid line represents the relation $\langle \sigma^2\rangle =\langle \delta^2 \rangle \cdot \tau_0^2$.}
\end{minipage}
\end{figure}

\end{document}

%% file: bibl.tex
%